\documentclass[%
 aip,
 jmp,%
 amsmath,amssymb,
 preprint,%
]{revtex4-1}

\usepackage{graphicx}
\usepackage{amscd}
\usepackage{amssymb}
\usepackage{amsmath}
\usepackage[squaren]{SIunits}
\usepackage{color}
\usepackage{subfigure}

\begin{document}
\preprint{AIP/123-QED}

\title[]{Circular photogalvanic effect induced by near-infrared radiation in InAs quantum wires patterned quasi two-dimensional electron system}
\author{Chongyun Jiang}
\author{Yonghai Chen}\email[]{yhchen@semi.ac.cn}
\author{Hui Ma}
\author{Jinling Yu}
\author{Yu Liu}

\affiliation{Key Laboratory of Semiconductor Material Science,
Institute of Semiconductors, Chinese Academy of Sciences, 100083 Beijing, China.}

\date{\today}

\begin{abstract}
In this work we investigated the InAs/InAlAs
quantum wires (QWRs) superlattice by optically exciting the structure with
near-infrared radiation. By varying
the helicity of the radiation at room temperature
we observed the circular photogalvanic
effect related to the $C_{2v}$ symmetry of the structure,
which could be attributed to the formation of
a quasi two-dimensional system underlying
in the vicinity of the QWRs pattern. The ratio of Rashba and Dresselhaus terms
shows an evolution of the spin-orbit interaction
in quasi two-dimensional structure with
the QWR layer deposition thickness.
\end{abstract}

\maketitle

In the last decade self-assembly grown nanostructures have attracted much research
interest\cite{Patan`eLevinPolimeniEavesMainHeniniHill2000,FossardHelmanFishmanJulienBraultGendryP'eronneAlexandrouSchachamBahirFinkman2004,WangChenWuLeiWangZeng2005,LeiChenJinYeWangXuWang2006,LeiChenWangHuangZhaoLiuXuJinZengWang2006}.
The self-assembled InAs quantum wires (QWRs) superlattice is one of the novel structures, which
are smaller in size compared with some other nano wires and thus exhibit quantum phenomena.
Precursors have done much work on these structures in the aspects of growing
technique\cite{LeiChenJinYeWangXuWang2006,LeiChenWangHuangZhaoLiuXuJinZengWang2006,WangChenWuLeiWangZeng2005},
microscopy and optical spectroscopy. However, InAs is of great interest due to its high mobility,
narrow band gap and large spin-orbit interaction.
Much investigation with respect to these properties could be enrolled.

Considering the alignment of the QWRs which is responsible for the anisotropy of optical and
electrical properties, we could invoke some techniques with respect to the symmetry.
Photogalvanic effect\cite{GanichevPrettl2003,Ivchenko2005,IvchenkoGanichev2008,Ivchenko2005a,LechnerGolubOlbrichStachelSchuhWegscheiderBel'kovGanichev2009,OlbrichIvchenkoRavashFeilDanilovAllerdingsWeissSchuhWegscheiderGanichev2009}
provides us such a way to investigate the symmetry of the structure and also
the spin or orbital behavior of the charge carriers. The phenomena of photogalvanic effect
is the generation of a direct electric current induced
by homogenous radiation in a homogenous sample.
Photogalvanic effect arises due to the absence of an inversion center in a crystal lattice.
If the radiation is circularly polarized, it is called circular photogalvanic effect (CPGE).
The CPGE current in a two-dimensional system of $C_{2v}$ symmetry can be written as\cite{GanichevPrettl2003,Ivchenko2005}
\begin{equation}
\label{eqn:jcy}
  j_{C,y}=e\tau_p \frac{\gamma_{yx}}{\hbar}\frac{\alpha (\lambda)I}{\hbar\omega}P_{c}\hat{e}_x,
\end{equation}
where $\tau_p$ is a typical momentum relaxation time,
$\alpha (\lambda)$ is the absorption coefficient, $I$ is the intensity of the radiation,
$P_{c}$ describes the helicity
of the radiation and $\hat{e}_x$ is the $x$ component of the
unit vector of the electric field amplitude.
$\gamma_{yx}$ is a second-rank pseudo-tensor
which correlates the helix excitation $P_{c}$ with
the direct current $j_{C,y}$. Further details of
CPGE in two-dimensional QWs can be find in Ref.
\onlinecite{GanichevPrettl2003}
and \onlinecite{GiglbergerGolubBel'kovDanilovSchuhGerlRohlfingStahlWegscheiderWeissPrettlGanichev2007}.
In a two-dimensional electron system, structure and bulk inversion asymmetry (SIA and BIA)
contribute to the zero-field spin splitting.
The SIA arises from the asymmetry in the grown direction of the structure,
whereas the BIA arises from the absence of an inversion center.
For a system of $C_{2v}$ symmetry, the relative magnitude of SIA and BIA can be
extracted via measuring the components of the current in
different directions\cite{GiglbergerGolubBel'kovDanilovSchuhGerlRohlfingStahlWegscheiderWeissPrettlGanichev2007},
which demonstrates the property in terms of the spin-orbit interaction
in the investigated low dimensional system.

\begin{figure}[htbp]
  \centering
  \includegraphics[width=7cm]{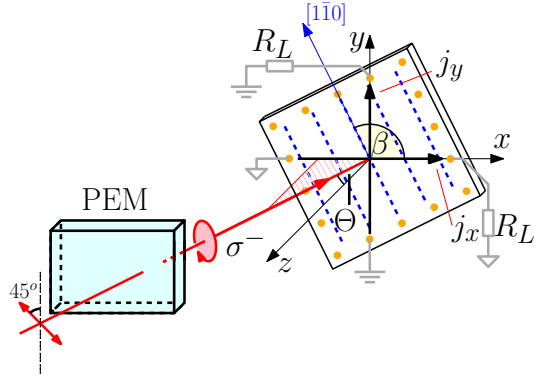}\\
  \caption{Experimental setup. The sample plane lies in the $x-y$ plane and the $x-z$
  plane is always the plane of incidence. $j_x$ and $j_y$ are the $x$ and $y$
  components of the current
  respectively.
  $\Theta$ is the angle of incidence between the incident light and the $z$ axes. $\beta$
  is the azimuth angle, which is between the orientation of QWRs and $x$ axes.
  Any component of the current $j_\alpha$ (The index $\alpha$ refers either $x$ or $y$
  with respect to the coordinate system)
  is calculated using the formula
$j_{\alpha}~=~\frac{\Delta V_{\alpha}}{R_L}$,
where $\Delta V_{\alpha}$ is the electric potential difference
between the two measured contacts,
$R_L$ is the load resistance with a value of 15 $\kilo\ohm$.}\label{fig:setup}
\end{figure}

Our QWRs samples are prepared by molecular-beam epitaxy (MBE)
technique on semi-insulating (001)-oriented InP substrates.
A $300 \nano\meter$ InAlAs buffer layer and
six periods of InAs-In$_{0.52}$Al$_{0.48}$As superlattice
are deposited. In each period of the InAs-In$_{0.52}$Al$_{0.48}$As superlattice
there is one QWR layer with a thickness of 2 $\mega\liter$, 4 $\mega\liter$ or
6 $\mega\liter$ respectively for three samples and one
unsymmetrical Si-doped InAlAs space layer.
The Si donors are doped within layer with a thickness of 2 $\nano\meter$, which is
sandwiched between a 4 nm and a 9 nm thick
undoped InAlAs layer.
The doping concentration is $1\times 10^{18}$ $\centi\meter^{-3}$.
An 80 nm thick InAlAs cap layer is grown on the most top.
We cut the samples into 10$\times$10 $\milli\meter^2$ in size and alloyed sixteen
Ohmic contacts equidistantly on the edges (see Fig. \ref{fig:setup}).
According to the TEM images of the same samples in
Ref. \onlinecite{LeiChenWangHuangZhaoLiuXuJinZengWang2006},
the QWRs are aligned along $[1\bar{1}0]$-direction.
The photoluminescence and photocurrent
spectroscopy results of the samples and detailed discussions
can also be found in Ref. \onlinecite{LeiChenWangHuangZhaoLiuXuJinZengWang2006},
which show that the sample with thicker QWR layers has a narrower
size distribution of the wires and less defects.

The experimental setup is illustrated in Fig. \ref{fig:setup}.
The near-infrared laser radiation for the optical excitation has
a wavelength of 1.06 $\micro\meter$
and a power of $\sim$ 1 Watt.
A photoelastic modulator (PEM) is employed to convert the incoming
linearly polarized light into a modulated circularly polarized
light with a fixed
modulating frequency at 50 $\kilo\hertz$.

According to the working principle of the PEM,
the electrical signals from the samples referenced to the base frequency
correspond to the circular polarization,
while those referenced to the second harmonic frequency correspond to the linear polarization.
The electrical signals are therefore measured using standard lock-in technique.
The total electric current is given by
\begin{equation}
\label{eqn:jtotal}
    j~=~j_{C(1f)}+j_{L(2f)}+j_{0},
\end{equation}
where $j_{C(1f)}$ is the circular polarization contributed current,
$j_{L(2f)}$ is the linear polarization contributed current and $j_{0}$
is the polarization independent current which can be extracted by invoking an
optical chopper.

We firstly investigate
the photoconductivities\cite{BerrymanLyonSegev1997,ChuZrennerBoehmAbstreiter2000,Patan`eLevinPolimeniEavesMainHeniniHill2000}
of the samples in different directions in order to take the anisotropic absorbance into
account. The photoconductivity reads
\begin{equation}
\label{eqn:mobility}
  \Delta \sigma~=~\Delta n e \mu _n = [\alpha(\lambda)gI\tau _n]e \mu _n,
\end{equation}
where $\Delta n$
is the photogenerated excess carrier density,
$\alpha(\lambda)$ is the absorption coefficient, $g$ is the generation rate
of the carrier pairs, $I$ is the intensity of light, $\tau _n$ is the
life time of the excess electrons,
$e$ is elementary charge and $\mu _n$ is the
mobility of the electrons. Since the mobility of holes is one order of magnitude
lower than the electrons, the contribution of holes to the photoconductivity is
omitted in Eq. \ref{eqn:mobility}.
The investigation of photoconductivity is done by applying a DC-bias
between the contacts and recording the current referenced to
the optical chopper on the load resistor.
Since the photogalvanic effect is
proportional to the absorbance and intensity of the
radiation and the charge carrier density (See Eq. \ref{eqn:jcy}),
we could normalized the photogalvanic voltage by the photoconductivity
so that the photogalvanic currents in different directions are comparable
without referring to the anisotropic optical excitation.

\begin{figure}[htbp]
  \centering
  \includegraphics[width=9.5cm]{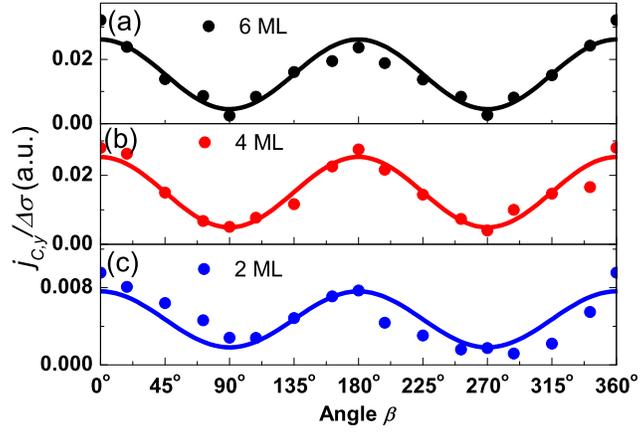}\\
  \caption{CPGE current $j_y$
  (See Fig. \ref{fig:setup} for the
  definition of the coordinate) with QWR layer thickness:
  (a) 2 $\mega\liter$, (b) 4 $\mega\liter$, (c) 6 $\mega\liter$
  as a function of azimuth angle $\beta$. The solid lines are the
  fitting curves. The angle of
  incidence is 30 $\degree$.}\label{fig:QWRs_3x2}
\end{figure}

The investigation of photogalvanic effect
is carried out at room temperature by varying the
azimuth angle $\beta$ of the incident light,
where the azimuth angle $\beta$ (See Fig. \ref{fig:setup}) is
the angle between the
plane of incidence and the alignment of the QWRs.
We vary $\beta$ by rotating the sample and measure the photocurrents
both perpendicular and parallel to the plane of incidence.
The CPGE currents are
shown in Fig. \ref{fig:QWRs_3x2}.
The experimental results
of the azimuth angle dependence of the CPGE current
can be well fitted by
\begin{eqnarray}
  \label{eqn:fitting}
  \begin{array}{l}
    j_{C,y}~=~ a_0+a_1\cos 2\beta,
  \end{array}
\end{eqnarray}
where $a_0$ and $a_1$ are fitting parameters.
In $[1\bar{1}0]$ direction, which is parallel to the
alignment of the QWRs, the current $j_{C,y}$ reaches minima.
In $[110]$ direction, the $j_{C,y}$ is maximum.
The results well demonstrate that the electrons are not confined in a single
wire but move in two dimensions.
We could extract the symmetric and anti-symmetric contributions
of the CPGE as a function of the azimuth angle. Compared with the two-dimensional
structure of $C_{2v}$ symmetry in the same framework as described in
Ref. \cite{GiglbergerGolubBel'kovDanilovSchuhGerlRohlfingStahlWegscheiderWeissPrettlGanichev2007},
we find that the current behaves similar to that in the two-dimensional system.
Thus, we suggest that the CPGE current is induced in a quasi two-dimensional structure
of $C_{2v}$ symmetry in the vicinity of the QWRs and subjected to an influence of the
QWRs pattern in terms of the anisotropic absorption. Therefore, we can also study
the structure and bulk inversion asymmetry (SIA and BIA)\cite{GiglbergerGolubBel'kovDanilovSchuhGerlRohlfingStahlWegscheiderWeissPrettlGanichev2007,PershinPiermarocchi2005}
in this quasi two-dimensional structure.

By taking the ratio of $a_0$ and $a_1$, we
can obtain the ratio of Rashba and Dresselhaus
terms\cite{GanichevBelchar39kovGolubIvchenkoSchneiderGiglbergerEromsBoeckBorghsWegscheiderWeissPrettl2004,Ivchenko2005a}
of the quasi two-dimensional system
\begin{equation}
  \label{eqn:ratio}
  \frac{R}{D}~=~\frac{a_0}{a_1},
\end{equation}
where $R$ is the Rashba coefficient and $D$ is the Dresselhaus
coefficient. The ratios $R/D$
in different samples are summarized in Table \ref{tab:RDratio}.
Since the samples
differ from one another only in the deposition thickness of
QWRs, the ratios of the SIA and BIA terms imply
an interplay of the QWRs structure and the quasi
two-dimensional system. As the thickness of the
QWR layer increasing, the $R/D$ ratios decrease, which indicates either a decrease of the SIA
or an increase of BIA. Since the samples only differ with one another in the size of the QWRs, whereas
the $\delta$-doping positions are the same,
the difference of the $R/D$ ratios comes from the coupling between the QWRs pattern and the quasi
two-dimensional structure.
The patterning effect imposed by the QWRs modulates the
zero-field spin splitting in terms of the ratio of Rashba and Dresselhaus terms.

\begin{figure}[htbp]
  \centering
  \includegraphics[width=8cm]{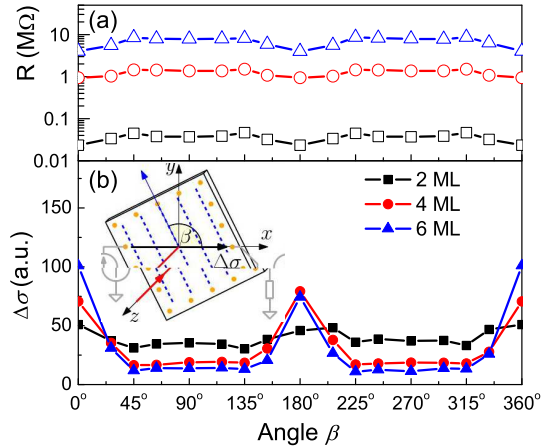}\\
  \caption{Resistance (a) and photoconductivity (b) of the samples as a
  function of the azimuth angle $\beta$. The squares, circles and
  triangles stand for the experimental data of the sample with 2~ML, 4~ML and
  6~ML thickness of QWR layer, respectively.
  }\label{fig:photoconductivity}
\end{figure}

The band gap of InAs bulk material is $0.354~\electronvolt$, which
is smaller than the photon energy of the $1.064~\micro\meter$
radiation ($1.19\electronvolt$).
The microscopic mechanism of optical transition
induced by the $1.064~\micro\meter$ radiation can be addressed to the
interband regime, where the electrons in the
valence band absorb the photon energy and transit to a higher level
in the conductance band.
According to Ref. \onlinecite{LeiChenWangHuangZhaoLiuXuJinZengWang2006},
the photoluminescence spectra show that the PL peak of
the structure with thicker InAs QWR layer has
a lower energy and smaller energy broadening, which indicates that
the size fluctuation decreases with increasing
InAs deposition thickness. Accordingly, the
electron could be more localized as the
increasing InAs deposition thickness.
However, since the excitation energy is
sufficiently high, photogenerated
carriers could be induced in some kind of
quasi two-dimensional structure in the vicinity of the QWRs, for instance,
the wetting layer between the buried InAs wires and the
In$_{0.52}$Al$_{0.48}$As barrier.
The electron wave function in a single
InAs wire couples with its neighbors.
Thus, the resistances do not differ identically in different directions (See Fig. \ref{fig:photoconductivity}(a)).
The electrons are free to move in two dimensions but subjected to the anisotropic influence
of the QWRs pattern in terms of the absorbance.

In summary, we observed the circular photogalvanic effect in the InAs QWRs patterned
quasi two-dimensional structure under 1.064 $\micro\meter$ near-infrared radiation. The current
exhibits anisotropy with respect to the alignment of the QWR pattern. The investigation of the
ratio of Rashba and Dresselhaus terms demonstrates a modulation of the spin-orbit interaction
in the quasi two-dimensional system by the QWRs pattern as the increasing deposition thickness.

\begin{acknowledgments}
This work is supported by the National Natural Science Foundation of China
(60625402, 60990313). C.Y. Jiang also thanks J.~Karch and S.D.~Ganichev in Regensburg
Terahertz Center for fruitful discussions.
\end{acknowledgments}

\nocite{*}

\begin{thebibliography}{16}%
\makeatletter
\providecommand \@ifxundefined [1]{%
 \@ifx{#1\undefined}
}%
\providecommand \@ifnum [1]{%
 \ifnum #1\expandafter \@firstoftwo
 \else \expandafter \@secondoftwo
 \fi
}%
\providecommand \@ifx [1]{%
 \ifx #1\expandafter \@firstoftwo
 \else \expandafter \@secondoftwo
 \fi
}%
\providecommand \natexlab [1]{#1}%
\providecommand \enquote  [1]{``#1''}%
\providecommand \bibnamefont  [1]{#1}%
\providecommand \bibfnamefont [1]{#1}%
\providecommand \citenamefont [1]{#1}%
\providecommand \href@noop [0]{\@secondoftwo}%
\providecommand \href [0]{\begingroup \@sanitize@url \@href}%
\providecommand \@href[1]{\@@startlink{#1}\@@href}%
\providecommand \@@href[1]{\endgroup#1\@@endlink}%
\providecommand \@sanitize@url [0]{\catcode `\\12\catcode `\$12\catcode
  `\&12\catcode `\#12\catcode `\^12\catcode `\_12\catcode `\%12\relax}%
\providecommand \@@startlink[1]{}%
\providecommand \@@endlink[0]{}%
\providecommand \url  [0]{\begingroup\@sanitize@url \@url }%
\providecommand \@url [1]{\endgroup\@href {#1}{\urlprefix }}%
\providecommand \urlprefix  [0]{URL }%
\providecommand \Eprint [0]{\href }%
\providecommand \doibase [0]{http://dx.doi.org/}%
\providecommand \selectlanguage [0]{\@gobble}%
\providecommand \bibinfo  [0]{\@secondoftwo}%
\providecommand \bibfield  [0]{\@secondoftwo}%
\providecommand \translation [1]{[#1]}%
\providecommand \BibitemOpen [0]{}%
\providecommand \bibitemStop [0]{}%
\providecommand \bibitemNoStop [0]{.\EOS\space}%
\providecommand \EOS [0]{\spacefactor3000\relax}%
\providecommand \BibitemShut  [1]{\csname bibitem#1\endcsname}%
\let\auto@bib@innerbib\@empty
\bibitem [{\citenamefont {Patan\`e}\ \emph {et~al.}(2000)\citenamefont
  {Patan\`e}, \citenamefont {Levin}, \citenamefont {Polimeni}, \citenamefont
  {Eaves}, \citenamefont {Main}, \citenamefont {Henini},\ and\ \citenamefont
  {Hill}}]{Patan`eLevinPolimeniEavesMainHeniniHill2000}%
  \BibitemOpen
  \bibfield  {author} {\bibinfo {author} {\bibfnamefont {A.}~\bibnamefont
  {Patan\`e}}, \bibinfo {author} {\bibfnamefont {A.}~\bibnamefont {Levin}},
  \bibinfo {author} {\bibfnamefont {A.}~\bibnamefont {Polimeni}}, \bibinfo
  {author} {\bibfnamefont {L.}~\bibnamefont {Eaves}}, \bibinfo {author}
  {\bibfnamefont {P.~C.}\ \bibnamefont {Main}}, \bibinfo {author}
  {\bibfnamefont {M.}~\bibnamefont {Henini}}, \ and\ \bibinfo {author}
  {\bibfnamefont {G.}~\bibnamefont {Hill}},\ }\bibfield  {title} {\enquote
  {\bibinfo {title} {Carrier thermalization within a disordered ensemble of
  self-assembled quantum dots},}\ }\href {\doibase 10.1103/PhysRevB.62.11084}
  {\bibfield  {journal} {\bibinfo  {journal} {Phys. Rev. B}\ }\textbf {\bibinfo
  {volume} {62}},\ \bibinfo {pages} {11084--11088} (\bibinfo {year}
  {2000})}\BibitemShut {NoStop}%
\bibitem [{\citenamefont {Fossard}\ \emph {et~al.}(2004)\citenamefont
  {Fossard}, \citenamefont {Helman}, \citenamefont {Fishman}, \citenamefont
  {Julien}, \citenamefont {Brault}, \citenamefont {Gendry}, \citenamefont
  {P\'eronne}, \citenamefont {Alexandrou}, \citenamefont {Schacham},
  \citenamefont {Bahir},\ and\ \citenamefont
  {Finkman}}]{FossardHelmanFishmanJulienBraultGendryP'eronneAlexandrouSchachamBahirFinkman2004}%
  \BibitemOpen
  \bibfield  {author} {\bibinfo {author} {\bibfnamefont {F.}~\bibnamefont
  {Fossard}}, \bibinfo {author} {\bibfnamefont {A.}~\bibnamefont {Helman}},
  \bibinfo {author} {\bibfnamefont {G.}~\bibnamefont {Fishman}}, \bibinfo
  {author} {\bibfnamefont {F.~H.}\ \bibnamefont {Julien}}, \bibinfo {author}
  {\bibfnamefont {J.}~\bibnamefont {Brault}}, \bibinfo {author} {\bibfnamefont
  {M.}~\bibnamefont {Gendry}}, \bibinfo {author} {\bibfnamefont
  {E.}~\bibnamefont {P\'eronne}}, \bibinfo {author} {\bibfnamefont
  {A.}~\bibnamefont {Alexandrou}}, \bibinfo {author} {\bibfnamefont {S.~E.}\
  \bibnamefont {Schacham}}, \bibinfo {author} {\bibfnamefont {G.}~\bibnamefont
  {Bahir}}, \ and\ \bibinfo {author} {\bibfnamefont {E.}~\bibnamefont
  {Finkman}},\ }\bibfield  {title} {\enquote {\bibinfo {title} {Spectroscopy of
  the electronic states in inas quantum dots grown on
  In$_{x}$Al$_{1-x}$As/InP(001)},}\ }\href {\doibase
  10.1103/PhysRevB.69.155333} {\bibfield  {journal} {\bibinfo  {journal} {Phys.
  Rev. B}\ }\textbf {\bibinfo {volume} {69}},\ \bibinfo {pages} {155333}
  (\bibinfo {year} {2004})}\BibitemShut {NoStop}%
\bibitem [{\citenamefont {Wang}\ \emph {et~al.}(2005)\citenamefont {Wang},
  \citenamefont {Chen}, \citenamefont {Wu}, \citenamefont {Lei}, \citenamefont
  {Wang},\ and\ \citenamefont {Zeng}}]{WangChenWuLeiWangZeng2005}%
  \BibitemOpen
  \bibfield  {author} {\bibinfo {author} {\bibfnamefont {Y.-L.}\ \bibnamefont
  {Wang}}, \bibinfo {author} {\bibfnamefont {Y.-H.}\ \bibnamefont {Chen}},
  \bibinfo {author} {\bibfnamefont {J.}~\bibnamefont {Wu}}, \bibinfo {author}
  {\bibfnamefont {W.}~\bibnamefont {Lei}}, \bibinfo {author} {\bibfnamefont
  {Z.-G.}\ \bibnamefont {Wang}}, \ and\ \bibinfo {author} {\bibfnamefont
  {Y.-P.}\ \bibnamefont {Zeng}},\ }\bibfield  {title} {\enquote {\bibinfo
  {title} {Structural and optical properties of
  InAs/In$_{0.52}$Al$_{0.48}$As self-assembled quantum wires on
  InP(001)},}\ }\href@noop {} {\bibfield  {journal} {\bibinfo  {journal}
  {Journal of Crystal Growth}\ }\textbf {\bibinfo {volume} {284}},\ \bibinfo
  {pages} {306--312} (\bibinfo {year} {2005})}\BibitemShut {NoStop}%
\bibitem [{\citenamefont {Lei}\ \emph {et~al.}(2006{\natexlab{a}})\citenamefont
  {Lei}, \citenamefont {Chen}, \citenamefont {Jin}, \citenamefont {Ye},
  \citenamefont {Wang}, \citenamefont {Xu},\ and\ \citenamefont
  {Wang}}]{LeiChenJinYeWangXuWang2006}%
  \BibitemOpen
  \bibfield  {author} {\bibinfo {author} {\bibfnamefont {W.}~\bibnamefont
  {Lei}}, \bibinfo {author} {\bibfnamefont {Y.~H.}\ \bibnamefont {Chen}},
  \bibinfo {author} {\bibfnamefont {P.}~\bibnamefont {Jin}}, \bibinfo {author}
  {\bibfnamefont {X.~L.}\ \bibnamefont {Ye}}, \bibinfo {author} {\bibfnamefont
  {Y.~L.}\ \bibnamefont {Wang}}, \bibinfo {author} {\bibfnamefont
  {B.}~\bibnamefont {Xu}}, \ and\ \bibinfo {author} {\bibfnamefont {Z.~G.}\
  \bibnamefont {Wang}},\ }\bibfield  {title} {\enquote {\bibinfo {title} {Shape
  and spatial correlation control of InAs-InAlAs-InP (001) nanostructure
  superlattices},}\ }\href@noop {} {\bibfield  {journal} {\bibinfo  {journal}
  {Applied Physics Letters}\ }\textbf {\bibinfo {volume} {88}},\ \bibinfo
  {pages} {063114} (\bibinfo {year} {2006}{\natexlab{a}})}\BibitemShut
  {NoStop}%
\bibitem [{\citenamefont {Lei}\ \emph {et~al.}(2006{\natexlab{b}})\citenamefont
  {Lei}, \citenamefont {Chen}, \citenamefont {Wang}, \citenamefont {Huang},
  \citenamefont {Zhao}, \citenamefont {Liu}, \citenamefont {Xu}, \citenamefont
  {Jin}, \citenamefont {Zeng},\ and\ \citenamefont
  {Wang}}]{LeiChenWangHuangZhaoLiuXuJinZengWang2006}%
  \BibitemOpen
  \bibfield  {author} {\bibinfo {author} {\bibfnamefont {W.}~\bibnamefont
  {Lei}}, \bibinfo {author} {\bibfnamefont {Y.~H.}\ \bibnamefont {Chen}},
  \bibinfo {author} {\bibfnamefont {Y.~L.}\ \bibnamefont {Wang}}, \bibinfo
  {author} {\bibfnamefont {X.~Q.}\ \bibnamefont {Huang}}, \bibinfo {author}
  {\bibfnamefont {C.}~\bibnamefont {Zhao}}, \bibinfo {author} {\bibfnamefont
  {J.~Q.}\ \bibnamefont {Liu}}, \bibinfo {author} {\bibfnamefont
  {B.}~\bibnamefont {Xu}}, \bibinfo {author} {\bibfnamefont {P.}~\bibnamefont
  {Jin}}, \bibinfo {author} {\bibfnamefont {Y.~P.}\ \bibnamefont {Zeng}}, \
  and\ \bibinfo {author} {\bibfnamefont {Z.~G.}\ \bibnamefont {Wang}},\
  }\bibfield  {title} {\enquote {\bibinfo {title} {Optical properties of
  self-assembled InAs/InAlAs/InP quantum wires with different
  inas deposited thickness},}\ }\href@noop {} {\bibfield  {journal} {\bibinfo
  {journal} {Journal of Crystal Growth}\ }\textbf {\bibinfo {volume} {286}},\
  \bibinfo {pages} {23--27} (\bibinfo {year} {2006}{\natexlab{b}})}\BibitemShut
  {NoStop}%
\bibitem [{\citenamefont {Ganichev}\ and\ \citenamefont
  {Prettl}(2003)}]{GanichevPrettl2003}%
  \BibitemOpen
  \bibfield  {author} {\bibinfo {author} {\bibfnamefont {S.}~\bibnamefont
  {Ganichev}}\ and\ \bibinfo {author} {\bibfnamefont {W.}~\bibnamefont
  {Prettl}},\ }\bibfield  {title} {\enquote {\bibinfo {title} {Spin
  photocurrents in quantum wells},}\ }\href {\doibase 692HU} {\bibfield
  {journal} {\bibinfo  {journal} {Journal of Physics-Condensed Matter}\
  }\textbf {\bibinfo {volume} {15}},\ \bibinfo {pages} {R935--R983} (\bibinfo
  {year} {2003})}\BibitemShut {NoStop}%
\bibitem [{\citenamefont {Ivchenko}(2005{\natexlab{a}})}]{Ivchenko2005}%
  \BibitemOpen
  \bibfield  {author} {\bibinfo {author} {\bibfnamefont {E.~L.}\ \bibnamefont
  {Ivchenko}},\ }\enquote {\bibinfo {title} {Circular photo-galvanic and
  spin-galvanic effects},}\ \ (\bibinfo  {publisher} {Springer},\ \bibinfo
  {year} {2005})\ Chap.\ \bibinfo {chapter} {658}, pp.\ \bibinfo {pages}
  {23--50}\BibitemShut {NoStop}%
\bibitem [{\citenamefont {Ivchenko}\ and\ \citenamefont
  {Ganichev}(2008)}]{IvchenkoGanichev2008}%
  \BibitemOpen
  \bibfield  {author} {\bibinfo {author} {\bibfnamefont {E.~L.}\ \bibnamefont
  {Ivchenko}}\ and\ \bibinfo {author} {\bibfnamefont {S.~D.}\ \bibnamefont
  {Ganichev}},\ }\enquote {\bibinfo {title} {Spin physics in semiconductors:
  Spin-photogalvanics},}\ \ (\bibinfo  {publisher} {Springer},\ \bibinfo {year}
  {2008})\ Chap.~\bibinfo {chapter} {9}, pp.\ \bibinfo {pages}
  {245--277}\BibitemShut {NoStop}%
\bibitem [{\citenamefont {Ivchenko}(2005{\natexlab{b}})}]{Ivchenko2005a}%
  \BibitemOpen
  \bibfield  {author} {\bibinfo {author} {\bibfnamefont {E.}~\bibnamefont
  {Ivchenko}},\ }\href@noop {} {\emph {\bibinfo {title} {Optical Spectroscopy
  of Semiconductor Nanostructures}}}\ (\bibinfo  {publisher} {Alpha Science
  Int., Harrow, UK},\ \bibinfo {year} {2005})\BibitemShut {NoStop}%
\bibitem [{\citenamefont {Lechner}\ \emph {et~al.}(2009)\citenamefont
  {Lechner}, \citenamefont {Golub}, \citenamefont {Olbrich}, \citenamefont
  {Stachel}, \citenamefont {Schuh}, \citenamefont {Wegscheider}, \citenamefont
  {Bel'kov},\ and\ \citenamefont
  {Ganichev}}]{LechnerGolubOlbrichStachelSchuhWegscheiderBel'kovGanichev2009}%
  \BibitemOpen
  \bibfield  {author} {\bibinfo {author} {\bibfnamefont {V.}~\bibnamefont
  {Lechner}}, \bibinfo {author} {\bibfnamefont {L.~E.}\ \bibnamefont {Golub}},
  \bibinfo {author} {\bibfnamefont {P.}~\bibnamefont {Olbrich}}, \bibinfo
  {author} {\bibfnamefont {S.}~\bibnamefont {Stachel}}, \bibinfo {author}
  {\bibfnamefont {D.}~\bibnamefont {Schuh}}, \bibinfo {author} {\bibfnamefont
  {W.}~\bibnamefont {Wegscheider}}, \bibinfo {author} {\bibfnamefont {V.~V.}\
  \bibnamefont {Bel'kov}}, \ and\ \bibinfo {author} {\bibfnamefont
  {S.}~\bibnamefont {Ganichev}},\ }\bibfield  {title} {\enquote {\bibinfo
  {title} {Tuning of structure inversion asymmetry by the delta-doping position
  in (001)-grown gaas quantum wells},}\ }\href@noop {} {\bibfield  {journal}
  {\bibinfo  {journal} {Appl. Phys. Lett.}\ }\textbf {\bibinfo {volume} {94}},\
  \bibinfo {pages} {242109} (\bibinfo {year} {2009})}\BibitemShut {NoStop}%
\bibitem [{\citenamefont {Olbrich}\ \emph {et~al.}(2009)\citenamefont
  {Olbrich}, \citenamefont {Ivchenko}, \citenamefont {Ravash}, \citenamefont
  {Feil}, \citenamefont {Danilov}, \citenamefont {Allerdings}, \citenamefont
  {Weiss}, \citenamefont {Schuh}, \citenamefont {Wegscheider},\ and\
  \citenamefont
  {Ganichev}}]{OlbrichIvchenkoRavashFeilDanilovAllerdingsWeissSchuhWegscheiderGanichev2009}%
  \BibitemOpen
  \bibfield  {author} {\bibinfo {author} {\bibfnamefont {P.}~\bibnamefont
  {Olbrich}}, \bibinfo {author} {\bibfnamefont {E.~L.}\ \bibnamefont
  {Ivchenko}}, \bibinfo {author} {\bibfnamefont {R.}~\bibnamefont {Ravash}},
  \bibinfo {author} {\bibfnamefont {T.}~\bibnamefont {Feil}}, \bibinfo {author}
  {\bibfnamefont {S.~D.}\ \bibnamefont {Danilov}}, \bibinfo {author}
  {\bibfnamefont {J.}~\bibnamefont {Allerdings}}, \bibinfo {author}
  {\bibfnamefont {D.}~\bibnamefont {Weiss}}, \bibinfo {author} {\bibfnamefont
  {D.}~\bibnamefont {Schuh}}, \bibinfo {author} {\bibfnamefont
  {W.}~\bibnamefont {Wegscheider}}, \ and\ \bibinfo {author} {\bibfnamefont
  {S.~D.}\ \bibnamefont {Ganichev}},\ }\bibfield  {title} {\enquote {\bibinfo
  {title} {Ratchet effects induced by terahertz radiation in heterostructures
  with a lateral periodic potential},}\ }\href {\doibase
  10.1103/PhysRevLett.103.090603} {\bibfield  {journal} {\bibinfo  {journal}
  {Phys. Rev. Lett.}\ }\textbf {\bibinfo {volume} {103}},\ \bibinfo {pages}
  {090603} (\bibinfo {year} {2009})}\BibitemShut {NoStop}%
\bibitem [{\citenamefont {Giglberger}\ \emph {et~al.}(2007)\citenamefont
  {Giglberger}, \citenamefont {Golub}, \citenamefont {Bel'kov}, \citenamefont
  {Danilov}, \citenamefont {Schuh}, \citenamefont {Gerl}, \citenamefont
  {Rohlfing}, \citenamefont {Stahl}, \citenamefont {Wegscheider}, \citenamefont
  {Weiss}, \citenamefont {Prettl},\ and\ \citenamefont
  {Ganichev}}]{GiglbergerGolubBel'kovDanilovSchuhGerlRohlfingStahlWegscheiderWeissPrettlGanichev2007}%
  \BibitemOpen
  \bibfield  {author} {\bibinfo {author} {\bibfnamefont {S.}~\bibnamefont
  {Giglberger}}, \bibinfo {author} {\bibfnamefont {L.~E.}\ \bibnamefont
  {Golub}}, \bibinfo {author} {\bibfnamefont {V.~V.}\ \bibnamefont {Bel'kov}},
  \bibinfo {author} {\bibfnamefont {S.~N.}\ \bibnamefont {Danilov}}, \bibinfo
  {author} {\bibfnamefont {D.}~\bibnamefont {Schuh}}, \bibinfo {author}
  {\bibfnamefont {C.}~\bibnamefont {Gerl}}, \bibinfo {author} {\bibfnamefont
  {F.}~\bibnamefont {Rohlfing}}, \bibinfo {author} {\bibfnamefont
  {J.}~\bibnamefont {Stahl}}, \bibinfo {author} {\bibfnamefont
  {W.}~\bibnamefont {Wegscheider}}, \bibinfo {author} {\bibfnamefont
  {D.}~\bibnamefont {Weiss}}, \bibinfo {author} {\bibfnamefont
  {W.}~\bibnamefont {Prettl}}, \ and\ \bibinfo {author} {\bibfnamefont {S.~D.}\
  \bibnamefont {Ganichev}},\ }\bibfield  {title} {\enquote {\bibinfo {title}
  {Rashba and dresselhaus spin splittings in semiconductor quantum wells
  measured by spin photocurrents},}\ }\href@noop {} {\bibfield  {journal}
  {\bibinfo  {journal} {Phys. Rev. B}\ }\textbf {\bibinfo {volume} {75}},\
  \bibinfo {pages} {035327} (\bibinfo {year} {2007})}\BibitemShut {NoStop}%
\bibitem [{\citenamefont {Berryman}, \citenamefont {Lyon},\ and\ \citenamefont
  {Segev}(1997)}]{BerrymanLyonSegev1997}%
  \BibitemOpen
  \bibfield  {author} {\bibinfo {author} {\bibfnamefont {K.~W.}\ \bibnamefont
  {Berryman}}, \bibinfo {author} {\bibfnamefont {S.~A.}\ \bibnamefont {Lyon}},
  \ and\ \bibinfo {author} {\bibfnamefont {M.}~\bibnamefont {Segev}},\
  }\bibfield  {title} {\enquote {\bibinfo {title} {Mid-infrared
  photoconductivity in inas quantum dots},}\ }\href@noop {} {\bibfield
  {journal} {\bibinfo  {journal} {Appl. Phys. Lett.}\ }\textbf {\bibinfo
  {volume} {70}},\ \bibinfo {pages} {1861} (\bibinfo {year}
  {1997})}\BibitemShut {NoStop}%
\bibitem [{\citenamefont {Chu}\ \emph {et~al.}(2000)\citenamefont {Chu},
  \citenamefont {Zrenner}, \citenamefont {B\"{o}hm},\ and\ \citenamefont
  {Abstreiter}}]{ChuZrennerBoehmAbstreiter2000}%
  \BibitemOpen
  \bibfield  {author} {\bibinfo {author} {\bibfnamefont {L.}~\bibnamefont
  {Chu}}, \bibinfo {author} {\bibfnamefont {A.}~\bibnamefont {Zrenner}},
  \bibinfo {author} {\bibfnamefont {G.}~\bibnamefont {B\"{o}hm}}, \ and\
  \bibinfo {author} {\bibfnamefont {G.}~\bibnamefont {Abstreiter}},\ }\bibfield
   {title} {\enquote {\bibinfo {title} {Lateral intersubband photocurrent
  spectroscopy on InAs/GaAs quantum dots},}\ }\href@noop {}
  {\bibfield  {journal} {\bibinfo  {journal} {Appl. Phys. Lett.}\ }\textbf
  {\bibinfo {volume} {76}},\ \bibinfo {pages} {1944} (\bibinfo {year}
  {2000})}\BibitemShut {NoStop}%
\bibitem [{\citenamefont {Pershin}\ and\ \citenamefont
  {Piermarocchi}(2005)}]{PershinPiermarocchi2005}%
  \BibitemOpen
  \bibfield  {author} {\bibinfo {author} {\bibfnamefont {Y.~V.}\ \bibnamefont
  {Pershin}}\ and\ \bibinfo {author} {\bibfnamefont {C.}~\bibnamefont
  {Piermarocchi}},\ }\bibfield  {title} {\enquote {\bibinfo {title} {Spin
  photovoltaic effect in quantum wires with rashba interaction},}\ }\href@noop
  {} {\bibfield  {journal} {\bibinfo  {journal} {Appl. Phys. Lett.}\ }\textbf
  {\bibinfo {volume} {86}},\ \bibinfo {pages} {212107} (\bibinfo {year}
  {2005})}\BibitemShut {NoStop}%
\bibitem [{\citenamefont {Ganichev}\ \emph {et~al.}(2004)\citenamefont
  {Ganichev}, \citenamefont {Bel\char39{}kov}, \citenamefont {Golub},
  \citenamefont {Ivchenko}, \citenamefont {Schneider}, \citenamefont
  {Giglberger}, \citenamefont {Eroms}, \citenamefont {Boeck}, \citenamefont
  {Borghs}, \citenamefont {Wegscheider}, \citenamefont {Weiss},\ and\
  \citenamefont
  {Prettl}}]{GanichevBelchar39kovGolubIvchenkoSchneiderGiglbergerEromsBoeckBorghsWegscheiderWeissPrettl2004}%
  \BibitemOpen
  \bibfield  {author} {\bibinfo {author} {\bibfnamefont {S.~D.}\ \bibnamefont
  {Ganichev}}, \bibinfo {author} {\bibfnamefont {V.~V.}\ \bibnamefont
  {Bel\char39{}kov}}, \bibinfo {author} {\bibfnamefont {L.~E.}\ \bibnamefont
  {Golub}}, \bibinfo {author} {\bibfnamefont {E.~L.}\ \bibnamefont {Ivchenko}},
  \bibinfo {author} {\bibfnamefont {P.}~\bibnamefont {Schneider}}, \bibinfo
  {author} {\bibfnamefont {S.}~\bibnamefont {Giglberger}}, \bibinfo {author}
  {\bibfnamefont {J.}~\bibnamefont {Eroms}}, \bibinfo {author} {\bibfnamefont
  {J.~D.}\ \bibnamefont {Boeck}}, \bibinfo {author} {\bibfnamefont
  {G.}~\bibnamefont {Borghs}}, \bibinfo {author} {\bibfnamefont
  {W.}~\bibnamefont {Wegscheider}}, \bibinfo {author} {\bibfnamefont
  {D.}~\bibnamefont {Weiss}}, \ and\ \bibinfo {author} {\bibfnamefont
  {W.}~\bibnamefont {Prettl}},\ }\bibfield  {title} {\enquote {\bibinfo {title}
  {Experimental separation of rashba and dresselhaus spin splittings in
  semiconductor quantum wells},}\ }\href {\doibase
  10.1103/PhysRevLett.92.256601} {\bibfield  {journal} {\bibinfo  {journal}
  {Phys. Rev. Lett.}\ }\textbf {\bibinfo {volume} {92}},\ \bibinfo {pages}
  {256601} (\bibinfo {year} {2004})}\BibitemShut {NoStop}%
\end{thebibliography}
%

\begin{table}[htbp]
  \centering
  \caption{Ratio of Rashba and Dresselhaus terms}\label{tab:RDratio}
  \begin{tabular}{lccc}
    \hline
    Thickness of a QWR layer (ML)  &2 &4 &6\\
    \hline
    Average width of a QWR (nm)	&11.8&	14.4&	16.7\\
    Average height of a QWR (nm)	&3.8&	4.7&	5.6\\
    Ratio of SIA and BIA $(R/D$)	&1.50&	1.29&	1.25\\	
    \hline
  \end{tabular}
\end{table}

\end{document}